# Transition from Casimir to van der Waals force between macroscopic bodies

G. Palasantzas[*], P. J. van Zwol, and J. Th. M. De Hosson

Department of Applied Physics, Materials Innovation Institute and Zernike Institute for Advanced Materials, University of Groningen, Nijenborgh 4, 9747 AG Groningen, the Netherlands.

**Abstract**

The transition of van der Waals to Casimir forces between macroscopic gold surfaces is investigated by Atomic Force Microscopy in the plane-sphere geometry. It was found that the transition appears to take place at separations ~10 % the plasma wavelength $\lambda_p$ for evaporated gold surfaces, which compares to theoretical predictions by incorporation of experimental optical data and roughness corrections. Moreover, the force data allow estimation of the Hamaker constant $A_H$ in the van der Waals regime, which is in good agreement with the Lifshitz theory predictions (even if roughness corrections are taken into account) and former surface force apparatus measurements.

Pacs numbers: 78.68.+m, 03.70.+k, 85.85.+j, 12.20.Fv

---

[*]Corresponding author: g.palasantzas@rug.nl

When the proximity between material objects, e.g., electrodes in micro/nanoelectromechanical system (MEMS/NEMS) [1-3], becomes of the order of a few microns down to nanometers, a regime is entered in which forces that are quantum mechanical in nature, namely, van der Waals (vdW) and Casimir forces, become operative [1-5]. In fact, at separations below *100 nm*, the Casimir force is very strong and becomes comparable to electrostatic forces corresponding to voltages in the range *0.1-1V* [1-3], whereas for separations below *10 nm* vdW forces dominate any attraction [1, 4-9]. In addition, from the fundamental point of view, precise measurements of forces from nano to micrometer length scales have attracted considerable interest in a search for hypothetical force fields beyond the Standard Model [10].

Furthermore, as it was discussed recently, the crossover between the short and long-distance force laws is quite similar to the crossover between vdW and Casimir-Polder forces for two atoms in vacuum [11]. This result obtained at short distances can be understood as the London interaction between plasmon excitations at the surface of each bulk mirror. Moreover, calculations of the Casimir/vdW force in terms of the Lifshitz theory using for the optical properties the Drude model yielded a transition from Casimir to vdW regime at separations (*d*) ~*10 %* the plasma wave length $\lambda_p$. Moreover, fits of the vdW force in the form $\sim A_H/d^2$ yielded a Hamaker constant value $A_H \approx (7-25) \times 10^{-20}$ *J* for gold (Au)-water-gold systems [13]. For Au-air-Au surfaces, studies by Tonck et al. [13] using the surface force apparatus (SFA) in the plane –sphere geometry with millimeter size spheres yielded a Hamaker constant of $A_H \approx 28 \times 10^{-20}$ *J* for separations *d>8. 5 nm* for Au coatings having roughness of *5-6 nm* peak-to-peak.

Precision measurements and theoretical descriptions of the Casimir/vdW force are nontrivial since the knowledge of optical properties of real films have to be taken into account carefully [14], and uncertainties in the separation distance due to roughness of real surfaces has also to be carefully considered (besides other calibration factors) [8, 15]. Therefore, we will investigate the transition of Casimir to vdW forces between real Au surfaces commonly obtained by vacuum deposition, and estimate the (non-retarded) Hamaker constant characterizing the strength of the vdW forces by incorporation of the real measured optical properties and roughness corrections.

The Casimir/vdW force is measured using the PicoForce AFM (http://www.veeco.com/), between a sphere with a diameter of *100 μm* and an rms surface roughness amplitude of *1.2 nm* (attached on a gold coated *240 μm* long cantilever with stiffness *k=4 N/m*), and an Au coated silicon plate. Both sphere and plate are coated with *100 nm* Au within the same vacuum evaporator. After Au deposition, the rms roughness of sphere and plate were measured by AFM (see Fig. 1) to be *1.8* and *1.3 (±0.2) nm*, respectively. Moreover, analysis of the sphere where contact is takes place was investigated by inverse imaging (Fig. 1b) [15]. Moreover, electrostatic fitting in the range of *1 to 4 μm* with voltages in the range ± *(3-4.5) V* yielded the cantilever stiffness $k$ and contact potential $V_0$ (≈*10±10 mV*) [15]. The contact separation due to roughness $d_0$ was derived from the top-to-bottom roughness of sphere and plate (from multiple scans at different places of both surfaces) added and divided by two yielding $d_0$=*7.5±1 nm*. From the error in $d_0$ of *1 nm* we estimate a relative error in the force at the smallest separations as $\Delta F/F \approx m\Delta d/d$. since experimentally $F \sim 1/d^m$ with $m \approx$*2-3*. Finally, the optical properties of the Au film on the plate were measured with an ellipsometry in the wavelength range *137 nm – 33 μm* [8, 14, 15] yielding the Drude parameters $w_p$=*7.9±0.2 eV* and $w_t$=*0.048±0.005 eV* (fitting the optical data in the infrared range).

After calibration, the Casimir/vdW force is measured and averaged using 40 force curves. Calibration and measurements were repeated at 20 different locations on the plane having in total an average of 800 curves to obtain the force as depicted in Fig. 2. The transition from Casimir to vdW regime is rather weak. At separations larger than *20 nm* the force follows the power law $F \sim d^{-m_c}$ with $m_c$=2.5 ±0.03. Indeed, for the sphere-plate geometry with perfect conductors the expected value is $m_c$=3, while deviations are mainly due to finite conductivity corrections for relatively smooth surfaces or sufficiently large separations [8, 15]. The force is given by $F_{theory}=(2\pi R/A)E_{pp,rough}$ with $E_{pp,rough} = E_{ppflat} + \delta E_{pp,rough}$ the Casimir/vdW energy for parallel plates calculated using Lifshitz's theory [16]. This theory yields for flat surfaces

$$E_{ppflat} = -\hbar A \sum_P \int [d^2k/4\pi^2] \int_0^\infty [d\Phi/2\pi] \ln[1 - r^P(k,\Phi)^2 e^{-2\kappa l}] \tag{1}$$

with A the plane area, r(Φ) the reflection coefficient, Φ the imaginary frequency of the electromagnetic wave, and p the index denoting the transverse electric and magnetic modes [16]. The roughness correction is given by $\delta E_{pp,rough} = \int [d^2k/4\pi^2] G(k)\sigma(k)$ [16], where σ(k) is the roughness spectrum [15, 17, 18], and G(k) a response function derived in [16].

In order to illustrate more clearly the rather smooth transition from the Casimir to vdW regime, and to estimate the non retarded Hamaker constant from the relation [13]

$$F_{vdW} \cong -\frac{A_H R}{6d^2}, \tag{2}$$

we plotted in Fig. 3a the force vs. $d^{-2}$. Note that Eq.(2) is an approximation of the more general equation derived by Hamaker (ignoring also nonspherical interactions within the interaction volume) [13] $F = -2A_H R^3 / 3d^2(d+2R)^2$ in the limit $d<<R$. Anyway, the slope of the linear fit in Fig. 3a yields $A_H \approx (29.4 \pm 0.6) \times 10^{-20}$ J for separations $d<18$ nm. If we consider the plasma wave length $\lambda_p = 2\pi c / \omega_p$ with $\omega_p = 7.9$ eV from the measured optical data, we obtain $\lambda_p \approx 155$ nm. Therefore, the vdW regime is probed for separations $d \leq 12\% \lambda_p$. The latter is in good agreement with theoretical predictions of $d \sim 10\% \lambda_p$ for the crossover from the Casimir to vdW regime by calculations considered the Drude model behavior for perfect Au films ($\omega_p = 9$ eV yielding $\lambda_p \approx 136$ nm) [12].

For completeness, we compare directly in Fig. 3 calculations using the Lifshitz theory [16] where we consider also the influence of surface roughness besides that of measured optical properties (Fig. 3b). Indeed, roughness corrections up to the lowest separation that we probe experimentally ($d \geq 12$ nm) can give a contribution of the order of 15 %, which is, however, within the accuracy of force measurements as error analysis indicated [15]. For this reason we display the straight line in between data with and without the roughness correction in Fig. 3b. In any case, theory shows a crossover at ~22 nm, while as Fig. 3a shows in detail experimentally, the transition occurs below 18 nm.

Finally, we will discuss the obtained Hamaker constant in comparison to other experimental studies. In fact, values for $A_H$ obtained between Au-Au surfaces in water as the medium in between gave as a highest reported value by Biggs and Mulvaney [13] of $A_H \approx 25 \times 10^{-20}$ J. These values were lower than the theoretically predicted value of $A_H \approx 40 \times 10^{-20}$ J [13]. Indeed, fits of the theory yield $A_H$ with and without roughness corrections (but including in both cases the measured optical data for the Au films [14]). From Fig. 3b we obtained $A_H$(no-roughness)$\approx 26 \times 10^{-20}$ J and $A_H$(roughness)$\approx 32 \times 10^{-20}$ J, while the experiment yielded the intermediate value $A_H \approx (29.4 \pm 0.6) \times 10^{-20}$ J, which is within the predictions. The obtained theory and experimental values compare to those by Tonck et al. [13] $A_H = (28 \pm 0.02) \times 10^{-20}$ J using the SFA approach, while optical characterization of their Au films was not accurate enough to allow reliable comparison with theory as was illustrated recently in [14]. Nonetheless, it is not clear why the Hamaker constant by Biggs and Mulvaney for the Au-water-Au system [13] was significantly close to that of the Au-air-Au system since the general notion is that that the presence of a medium reduces $A_H$.

In conclusion, we investigated the transition of van der Waals to Casimir forces between macroscopic gold surfaces, and estimated the associated Hamaker constant of the vdW interactions between real Au surfaces in air. The analysis took into account the measured optical data within a wide range necessary for the theory description of these forces, as well roughness contributions. In fact, the obtained Hamaker constant in the vdW regime, which is in agreement with predictions based on the Lifshitz theory, is comparable to measurements obtained by SFA in a former study [13]. Note that at smaller separations the roughness corrections increase drastically but the scattering theory approach up to second order in roughness amplitude [16] is no longer valid to allow reliable estimation of the effects of the surface morphology. Below *12 nm* the cantilever jumps to contact took place due to attractive capillary forces from the water layer present on the surface under ambient conditions preventing further measurements at shorter ranges.

**Acknowledgements:** The research was carried out under project number MC3.05242 in the framework of the Strategic Research programme of the Materials Innovation Institute (M2I). Financial support from the M2I is gratefully acknowledged.

**Figure captions**

**Figure 1 (a)** AFM topography with scan size 1 μm, and an associated height profile indicative of the roughness variations. **(b)** Inverse imaging of the sphere area (after Au deposition) around which contact with the surface occurs during force measurement, and an associated height profile indicative of the roughness height fluctuations.

**Figure 2** Force vs. separation d from average of 800 independent measurements (also averaged for two different spheres) with the power laws indicted for van der Waals and Casimir regimes. The arrow indicates qualitatively the transition regime approximately below *18 nm*.

**Figure 3 (a)** Force vs. $d^{-2}$ curves to illustrate the transition from vdW to Casimir regime. The slope of the linear fit yields the non retarded Hamaker constant $A_H$. **(b)** Calculations of force vs. $d^{-2}$ by incorporating the measured optical data for the Au films and roughness contribution. The force curve without roughness correction is given by (○), and that including the roughness contribution by (▲). The lower points (●) indicate the roughness correction for clarity purposes. For the roughness parameters of sphere and plane we used $w_{sphere}$ = *1.8 nm*, $w_{plane}$ = *1.3 nm,* lateral correlation lengths $\xi_{sphere, plane}$ = *20 nm,* and roughness exponent $H_{sphere, plane}$ = *0.9* [17]. The roughness parameters (w, ξ, H) for sphere and plane were determined by AFM measurement of the height correlation function $H(r) = <[h(r)-h(0)]^2>$ with <...> the ensemble average over multiple surface scans. The arrows indicate qualitatively the transition regime.

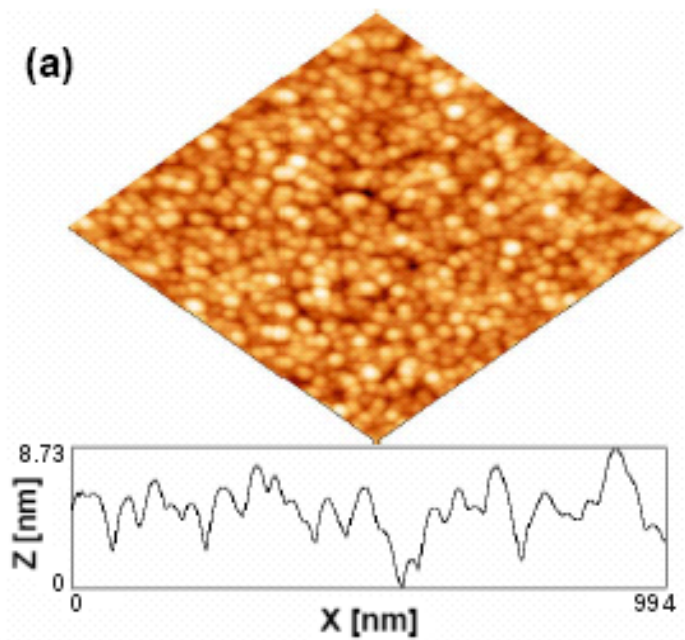
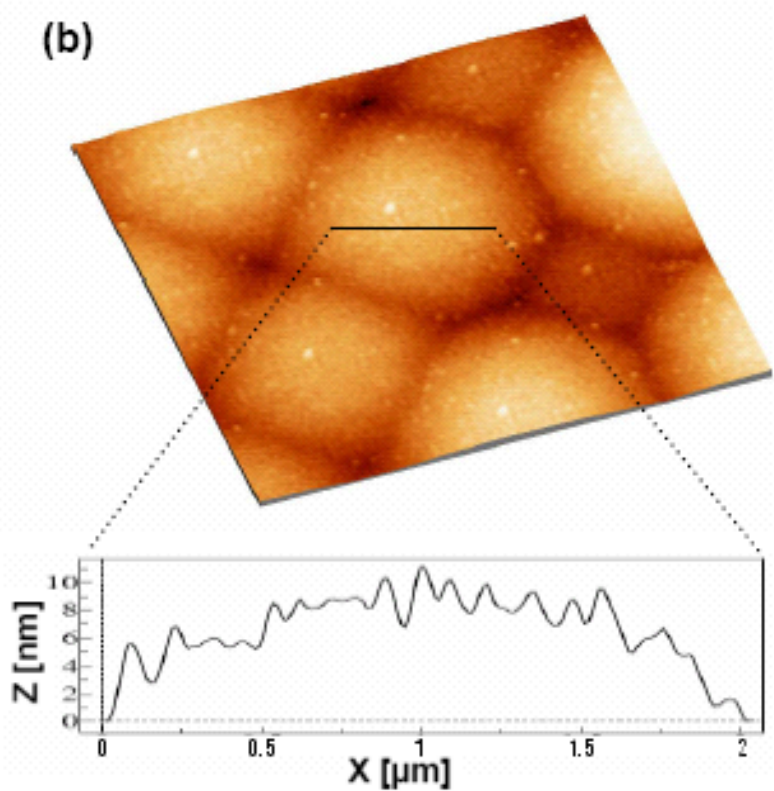

Figure 1

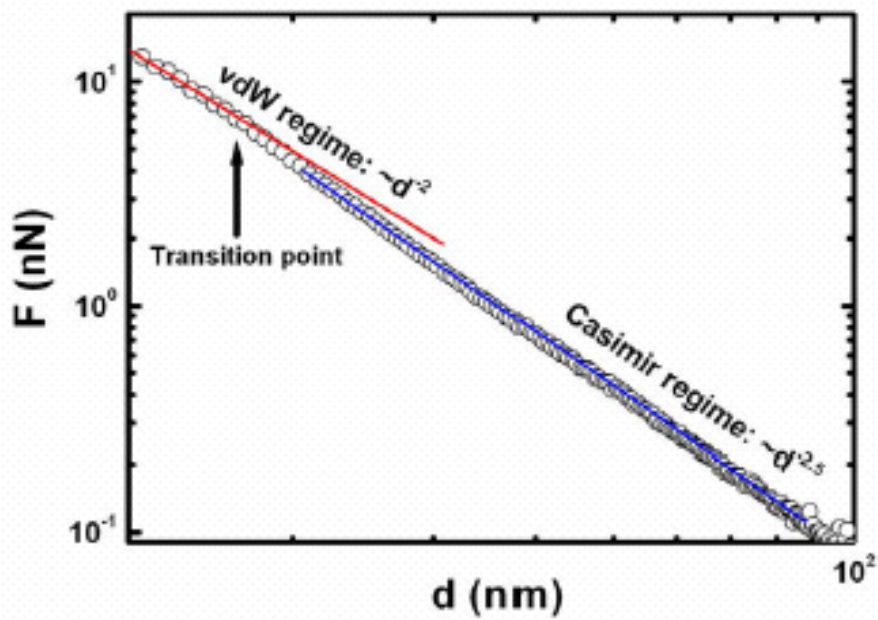

Figure 2

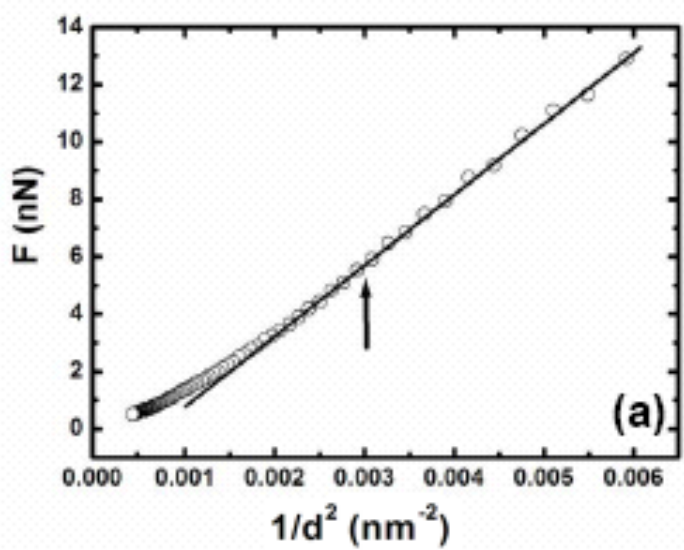

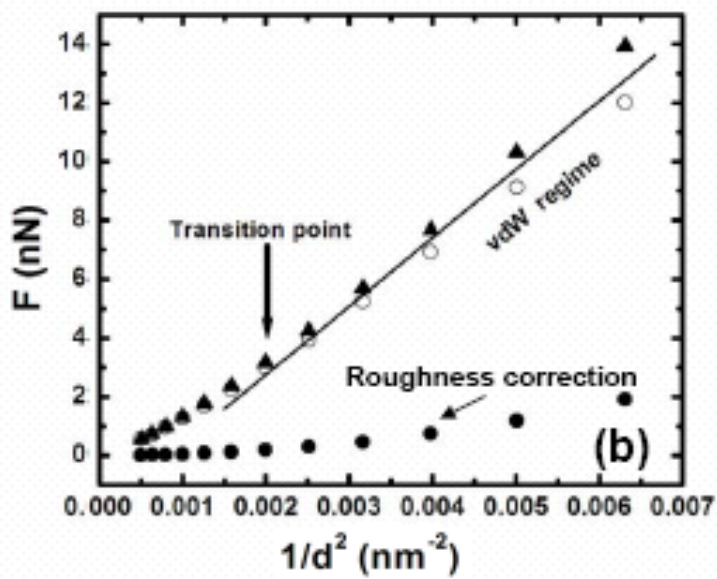

Figure 3